\documentclass[english,aps,reprint,twocolumn,superscriptaddress,longbibliography,showkeys,showpacs]{revtex4-2}
\usepackage[english]{babel}

\usepackage{amsmath,amssymb}
\usepackage{graphicx,xcolor}
\usepackage[unicode,bookmarksnumbered,breaklinks,colorlinks,linktocpage,
citecolor=blue,linkcolor=darkred,urlcolor=darkblue]{hyperref}
\usepackage{bm}
\usepackage{hyperref}

\definecolor{darkblue}{rgb}{0.0, 0.0, 0.55}
\definecolor{darkgreen}{rgb}{0.0, 0.2, 0.13}
\definecolor{darkred}{rgb}{0.55, 0.0, 0.0}

\begin{document}

\title{Manipulation of electromagnetic wave propagation in quantum-spin-chain medium}

\author{Taras Krokhmalskii} 
\affiliation{Yukhnovskii Institute for Condensed Matter Physics 
	of the National Academy of Sciences of Ukraine,\\
	Svientsitskii Street 1, 79011 L'viv, Ukraine}

\author{Taras Verkholyak}
\affiliation{Yukhnovskii Institute for Condensed Matter Physics 
	of the National Academy of Sciences of Ukraine,\\
	Svientsitskii Street 1, 79011 L'viv, Ukraine}
\affiliation{Professor Ivan Vakarchuk Department for Theoretical Physics,
	Ivan Franko National University of L’viv,\\
	Drahomanov Street 12, 79005 L’viv, Ukraine}

\author{Ostap Baran}
\affiliation{Yukhnovskii Institute for Condensed Matter Physics 
	of the National Academy of Sciences of Ukraine,\\
	Svientsitskii Street 1, 79011 L'viv, Ukraine}

\author{Dmytro Yaremchuk}
\affiliation{Yukhnovskii Institute for Condensed Matter Physics 
	of the National Academy of Sciences of Ukraine,\\
	Svientsitskii Street 1, 79011 L'viv, Ukraine}
\affiliation{Institute of Applied Mathematics and Fundamental Sciences,
	L’viv Polytechnic National University, 79013 L’viv, Ukraine}
	
\author{Taras Hutak}
\affiliation{Yukhnovskii Institute for Condensed Matter Physics 
	of the National Academy of Sciences of Ukraine,\\
	Svientsitskii Street 1, 79011 L'viv, Ukraine}

\author{Oleg Derzhko}
\email[Corresponding author: ]{derzhko@icmp.lviv.ua}
\affiliation{Yukhnovskii Institute for Condensed Matter Physics 
	of the National Academy of Sciences of Ukraine,\\
	Svientsitskii Street 1, 79011 L'viv, Ukraine}

\date{\today}

\begin{abstract}
We consider a simple model of one-dimensional magnetic crystal and examine the propagation of an electromagnetic wave through such a medium. Calculating the dispersion relation ${\bm k}(\omega)$ allows us to illustrate how the spread of the electromagnetic wave can be controlled by an external magnetic field. Our rigorous calculations should be useful for more realistic (and less tractable mathematically) models of magnetic media.
\end{abstract}


 
\maketitle

We dedicate this work to the memory of Johannes Richter. T.~K., T.~V., and O.~D. wrote a couple of papers on free-fermion spin chains in collaboration with Johannes, and they remember those times with great pleasure.

\section{Introduction}
\label{s1}

In quantum many-body systems research, a study of real-time equilibrium dynamics remains academically interesting and potentially practically useful. On one hand, such space- and time-dependent correlations yield information not only on the energy eigenvalues but also on the transitions between eigenstates; choosing certain dynamical variables allows to design theoretical probes on a specific purpose. On the other hand, there are plenty of measurement tools yielding experimental probes; for example, neutron scattering or electron spin resonance to name just a few. Another example, which is in focus of our investigation, concerns a response of the system to oscillating electric and magnetic fields, which occurs as electromagnetic wave propagates through the system. A weak interaction with the matter, hidden in certain equilibrium dynamical correlations, results in some changes of the electromagnetic wave characteristics. Understanding details of the electromagnetic wave propagation allows for their control and use for practical applications.

In the present study we focus on a simple quantum-spin-chain medium, for which theoretical analysis can be developed rather comprehensively. Namely, we consider a three-dimensional crystalline compound with quasi-one-dimensional exchange coupling between magnetic ions carrying spin $1/2$ \cite{Fazekas1999}. Moreover, this coupling is the isotropic $XY$ interaction and each ion experiences an external field directed along $z$ axis. Thus, we face a spin-1/2 isotropic $XY$ chain in a transverse field \cite{Lieb1961,Katsura1962,Katsura1963}. The main worth of this model is its exact solvability: All relevant quantities can be calculated rigorously and examined in great detail \cite{Lieb1961,Katsura1962,Katsura1963,Niemeijer1967,McCoy1971,Capel1977} (see also references in \cite{Jedrzejewski2008,Derzhko2008}). Moreover, such calculations may serve as a benchmark for more realistic and complicated cases which are not amenable to rigorous solutions.  Our goal is to study electromagnetic wave propagation in the considered magnetic crystal. As a result, we shall be able to discuss magnetic field control of electromagnetic wave propagation in the medium under examination.

It is worthwhile to discuss already at the beginning the scales relevant for this problem. The energy scale is dictated by the exchange coupling between magnetic ions, which may vary, say, in the region between $J/k_{\rm B}=10$~K and $J/k_{\rm B}=100$~K (here $k_{\rm B}$ is the Boltzmann constant). Equating $J$ to $h\nu$ (here $h$ is the Planck constant) we get $\nu\approx0.208{-}2.084$~terahertz that according to the relation $\lambda=c/\nu$ corresponds to the wavelength $\lambda\approx 1.44{-}0.14$~mm (here $c$ is the speed of light in vacuum). Furthermore, since the interatomic distance is about a nanometer, i.e., it is several orders smaller than $\lambda$, the spin-lattice model weakly couples, as a matter of fact, to spatially uniform magnetic field of the  electromagnetic wave. In other words, only the $\kappa\to 0$ Fourier mode of the spin-lattice system is relevant for the electromagnetic wave propagation under investigation. Furthermore, equating $J$ to $\mu_{\rm B}{\cal B}$ (here $\mu_{\rm B}\approx9.274\times 10^{-24}$~J/T (or $\approx 0.672$~K/T) is the Bohr magneton) we find for relevant fields $\approx 15{-}150$~T.
In our theoretical calculations below we often set $\hbar=h/(2\pi)=1$ and $k_{\rm B}=1$ for brevity.

The rest of this paper is organized as follows. In Sections~\ref{s2} and \ref{s3}, we briefly introduce notations for the considered free-fermion spin-chain model and recall the standard treatment of electromagnetic wave propagation through a medium. Section~\ref{s4} concerns dynamic susceptibilities, which enter the formulas for the  dispersion relation and determine the electromagnetic wave propagation. We explain in some detail how to obtain many-fermion dynamic susceptibilities. Finally, we discuss and summarize our findings in Section~\ref{s5}. 
Preliminary notes concerning this study were announced in Ref.~\cite{Baran2025}.
 
\section{Free-fermion spin-chain model}
\label{s2}

\begin{figure}
\includegraphics[width=\columnwidth]{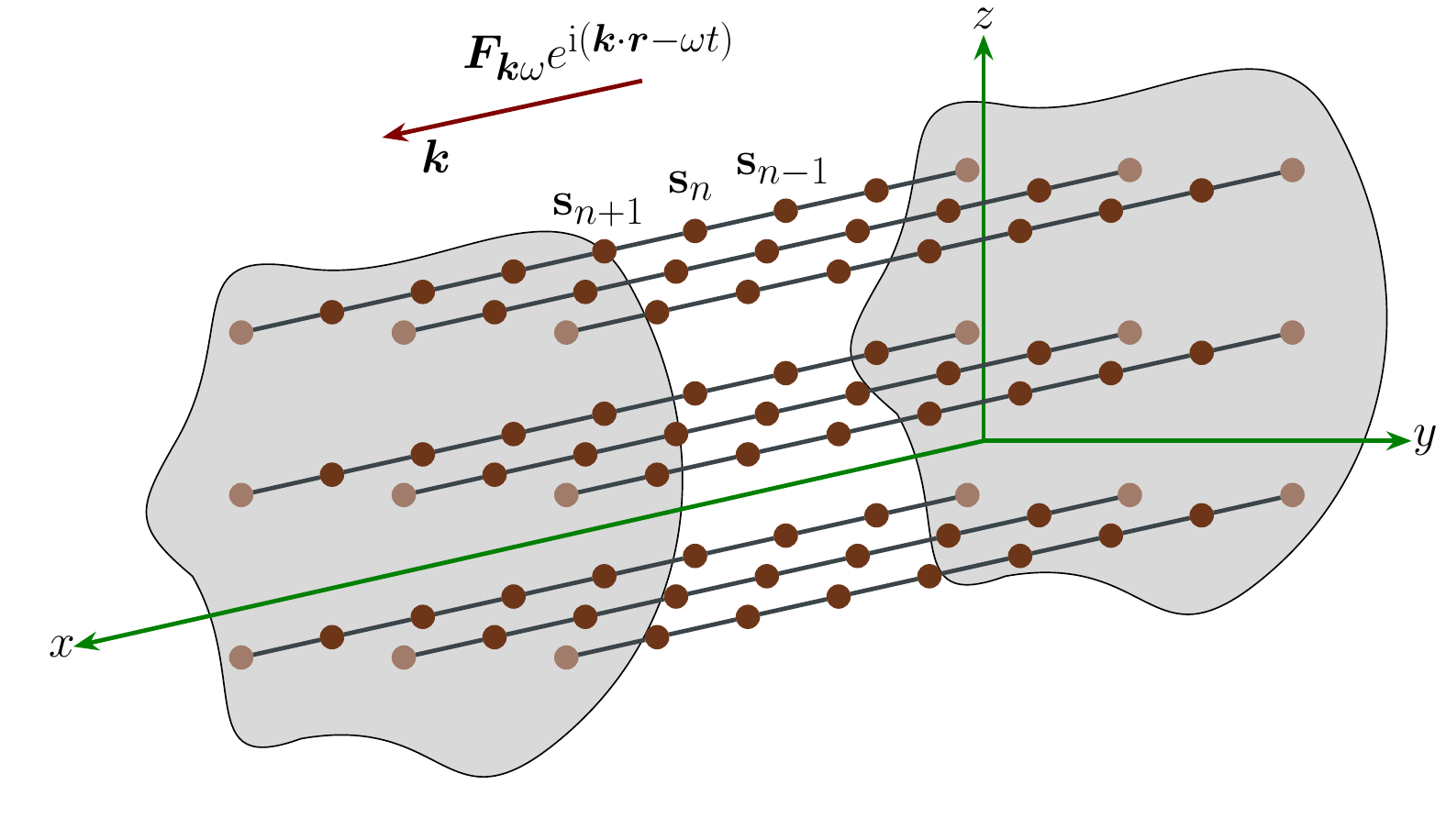}
\caption{Schematic drawing of electromagnetic wave propagation in quantum-spin-chain medium. An electromagnetic wave propagates along the $x$ direction, i.e., ${\bm k}{=}(k^x,0,0)$. Case 1: ${\bm E}_{{\bm k}\omega}{=}(0,E^y,0)$, ${\bm B}_{{\bm k}\omega}{=}(0,0,B^z)$. Case 2: ${\bm E}_{{\bm k}\omega}{=}(0,0,E^z)$, ${\bm B}_{{\bm k}\omega}{=}(0,B^y,0)$.}
\label{f1}
\end{figure}

In what follows we consider an $N$-site spin-1/2 chain with the following Hamiltonian:
\begin{eqnarray}
\label{01}
H{=}\sum_n J\left(s_n^xs_{n{+}1}^x{+}s_n^ys_{n{+}1}^y\right) {-}{\cal B}\sum_n m^z_n
\nonumber\\
=\sum_n\frac{J}{2}\left(s_n^+s_{n{+}1}^-{+}s_n^-s_{n{+}1}^+\right) 
{-}{\cal B}m_0\sum_n\left(s_{n}^+s_{n}^-{-}\frac{1}{2}\right).
\end{eqnarray}
Here, $J$ is the nearest-neighbor isotropic $XY$ interaction (below we often set $\vert J\vert=1$ to fix the units), ${\bm B}=(0,0,{\cal B})$, ${\cal B}\ge 0$ is an external field, whereas $m^z_n=m_0s_n^z$ and $m_0$ stands for the magnetic dipole moment (usually $m_0{=}g\mu_{\rm B}s$, $s{=}1/2$, and $g$ is around 2). Besides, we have introduced the on-site operators $s^{\pm}=s^x{\pm}{\rm i}s^y$ and $s^z=s^+s^-{-}1/2$. In the present study we focus on the antiferromagnetic interaction, i.e., $J>0$. In addition, we assume that the axes in the real and spin spaces coincide, and the chain runs along the $x$ direction, see Fig.~\ref{f1}.

Various properties of the introduced model can be analyzed thoroughly, without making any approximation, by using the standard Jordan-Wigner fermionization method \cite{Lieb1961,Katsura1962,Katsura1963,Niemeijer1967,McCoy1971,Capel1977,Jedrzejewski2008,Derzhko2008}. In brief, making use of the Jordan-Wigner transformation we introduce Fermi operators 
\begin{eqnarray}
\label{02}
c^{\dagger}_1=s^+_1,
\;\;\;
c^{\dagger}_n=(-2s^z_1)\ldots(-2s^z_{n-1})s^+_n,
\nonumber\\
c_1=s^-_1,
\;\;\;
c_n=(-2s^z_1)\ldots(-2s^z_{n-1})s^-_n,
\end{eqnarray}
in terms of which the spin Hamiltonian becomes a bilinear Fermi form: $H=\sum_{n,m}c_n^{\dagger}H_{nm}c_m+{\cal B}m_0N/2$. This form can be diagonalized after determining the eigenvalues and eigenvectors of $N{\times}N$ matrix $\left(H_{nm}\right)$. Now, inverting transformation (\ref{02}) results in the following formulas:
\begin{eqnarray}
\label{03}
s_n^x=\frac{1}{2}\varphi_1^+\varphi_1^-{\ldots}\varphi_{n-1}^+\varphi_{n-1}^-\varphi_n^+,
\nonumber\\
s_n^y=\frac{1}{2{\rm i}}\varphi_1^+\varphi_1^-{\ldots}\varphi_{n-1}^+\varphi_{n-1}^-\varphi_n^-,
\nonumber\\
s_n^z=-\frac{1}{2}\varphi_{n}^+\varphi_{n}^-,
\end{eqnarray}
where we have introduced the on-site operators $\varphi^{\pm}=c^{\dagger}\pm c$. 
It is clear now that a calculation of the thermodynamic average of a spin operator product $\langle s^{\alpha_1}_{j_1}(t_1){\ldots}s^{\alpha_m}_{j_m}(t_m)\rangle$, $s(t){=}e^{{\rm i}tH}se^{{-}{\rm i}tH}$, $\langle({\ldots})\rangle{=}{\rm Tr}(e^{{-}H/T}({\ldots}))/{\rm Tr}e^{{-}H/T}$, reduces to a straightforward  application of the Wick theorem, see below.

For periodic boundary condition imposed, after the Fourier transformation, $c_\kappa=\sum_n e^{{\rm i}\kappa n}c_n/\sqrt{N}$, Eq.~(\ref{01}) can be cast into the Hamiltonian of $N$ Fermi oscillators, 
\begin{eqnarray}
\label{04}
H=\sum_\kappa\epsilon_\kappa\left(c^{\dagger}_\kappa c_\kappa-\frac{1}{2}\right),
\;\;\; 
\epsilon_\kappa=J\cos\kappa-{\cal B}m_0,
\end{eqnarray}
$\kappa=2\pi z/N$, $z=-N/2,-N/2+1,\ldots,N/2-1$ ($N\to\infty$ is even). 
The ground-state phase diagram of model (\ref{01}) is quite simple: Until $0\le {\cal B}m_0<\vert J\vert$ the model is a spin liquid, whereas for $\vert J\vert<{\cal B}m_0$ it is a saturated paramagnet. While in the spin-liquid phase interaction is important, it becomes irrelevant in the paramagnetic phase which is similar to a collection of noninteracting spins. At any finite temperature the difference between ground-state phases disappears: Sharp boundaries become blurred.

The case of spin-1/2 isotropic Heisenberg chain results in interacting fermions. Indeed, Hamiltonian (\ref{01}) with additional interaction $J\sum_n s^z_ns^z_{n+1}$ after fermionization contains the term $J\sum_n(c_n^\dagger c_n{-}1/2)(c_{n{+}1}^\dagger c_{n{+}1}{-}1/2)$ and hence can be reduced to free fermions only after making a Hartree-Fock approximation \cite{Bulaevskii1963,Yamamoto2005}.

In our study, we use microscopic model (\ref{01}) for calculation of frequency-dependent susceptibilities, which enter the dispersion relation for the electromagnetic wave propagation, see Section~\ref{s3}. The advantage of the model under consideration is the possibility to calculate the needed dynamic quantities in the fermionic picture exactly, without making any approximate simplification, see Section~\ref{s4}. Other models would require some (approximate) approach (e.g., Green's functions or numerics) for calculation of dynamical correlations.

\section{Electromagnetic wave in quantum-spin-chain medium}
\label{s3}

Electromagnetic wave in a medium interacts with the matter of the medium that results, e.g., in the change of the electromagnetic wave speed. Theoretical analysis of the electromagnetic wave propagation in media uses Maxwell's equations in matter, which relate the fields ${\bm E}$, ${\bm D}$, ${\bm B}$, ${\bm H}$ and free charges $\rho$ and free currents ${\bm j}$ to each other:
\begin{eqnarray}
\label{05}
{\rm div}{\bm D}=4\pi\rho,
\;\;\;
{\rm div}{\bm B}=0,
\nonumber\\
{\rm rot}{\bm E}=-\frac{1}{c}\frac{\partial{\bm B}}{\partial t},
\;\;\;
{\rm rot}{\bm H}=\frac{4\pi}{c}{\bm j} +\frac{1}{c}\frac{\partial{\bm D}}{\partial t}
\end{eqnarray}
(Gaussian system); besides, these equations must be supplemented by the constitutive relations like ${\bm D}={\bm E}+4\pi{\bm P}$, ${\bm H}={\bm B}-4\pi{\bm M}$, where the polarization ${\bm P}$ (electric dipole moment per unit volume) and the magnetization ${\bm M}$ (magnetic dipole moment per unit volume) describe the material's electric and magnetic response to fields ${\bm E}$ and ${\bm H}$. Here, all quantities depend on position ${\bm r}$ and time $t$, the Cartesian components of fields are denoted as, e.g., $D_\alpha$, $\alpha=x,y,z$ and so on. Below, we recall standard reasonings keeping in mind the microscopic model of Sec.~\ref{s2}.

Since we consider a monochromatic uniform plane wave that propagates through a nonconducting, nonpolarizable, but magnetically active medium, it is convenient to present any field ${\bm F({\bm r},t)}$ in a complex form, that is, ${\bm F}_{{\bm k}\omega} \exp[{\rm i}({\bm k}{\cdot}{\bm r}{-}\omega t)]$, where ${\bm k}$ is the wave vector and $\omega$ is the frequency. Then Maxwell's equations (\ref{05}) supplemented by the constitutive relations read
\begin{eqnarray}
\label{06}
{\bm k}\cdot{\bm E}_{{\bm k}\omega}=0,
\;\;\; 
{\bm k}\cdot{\bm B}_{{\bm k}\omega}=0,
\nonumber\\
{\bm k}\times {\bm E}_{{\bm k}\omega}=\frac{\omega}{c}{\bm B}_{{\bm k}\omega},
\;\;\;
{\bm k}\times {\bm H}_{{\bm k}\omega}=-\frac{\omega}{c}{\bm E}_{{\bm k}\omega},
\nonumber\\
{\bm H}_{{\bm k}\omega}={\bm B}_{{\bm k}\omega}-4\pi{\bm M}_{{\bm k}\omega},
\end{eqnarray}
and for a magnetic medium
\begin{eqnarray}
\label{07}
M_{{\bm k}\omega}^{\alpha}=\chi_{\alpha\beta}({\bm k},\omega)H^{\beta}_{{\bm k}\omega}.
\end{eqnarray}
Here, $\chi_{\alpha\beta}({\bm k},\omega)=\chi_{\alpha\beta}$ stands for the wavevector- and frequen\-cy-dependent magnetic susceptibility tensor of the medium under consideration.

Bearing in mind the model of Sec.~\ref{s2}, we focus on the case when ${\bm k}=(k^x,0,0)$: Electromagnetic wave propagates along linear array of nearest-neighbor coupled magnetic ions, see Fig.~\ref{f1}. Moreover, we may consider separately the following two cases:
\begin{itemize}
\item 
${\bm E}_{{\bm k}\omega}=(0,E_{{\bm k}\omega}^y,0)$, $E_{{\bm k}\omega}^y=E^y$ and ${\bm B}_{{\bm k}\omega}=(0,0,B_{{\bm k}\omega}^z)$, $B_{{\bm k}\omega}^z=B^z$, that is, a plane wave travels along the $x$ (i.e., chain) direction with the electric/magnetic field oscillating along the $y/z$ direction 
(case 1)
\end{itemize}
and
\begin{itemize}
\item 
${\bm E}_{{\bm k}\omega}=(0,0,E_{{\bm k}\omega}^z)$, $E_{{\bm k}\omega}^z=E^z$ and ${\bm B}_{{\bm k}\omega}=(0,B_{{\bm k}\omega}^y,0)$, $B_{{\bm k}\omega}^y=B^y$, that is, a plane wave travels along the $x$ (i.e., chain) direction with the electric/magnetic field oscillating along the $z/y$ direction
(case 2). 
\end{itemize}

For the case 1,
according to Eqs.~(\ref{06}) and (\ref{07}), $E^y=E_{{\bm k}\omega}^y$ and $H^z=B^z-4\pi M^z$, $B^z=B_{{\bm k}\omega}^z$ satisfy
\begin{eqnarray}
\label{08}
\left(
\begin{array}{cc}
k^x  & -\mu_{zz} \frac{\omega}{c} \\
{-}\frac{\omega}{c} & k^x 
\end{array}
\right)
\left(
\begin{array}{c}
E^y \\
H^z
\end{array}
\right){=}0,
\end{eqnarray}
where we have introduced the wavevector- and frequency-dependent magnetic permeability $\mu_{\alpha\alpha}=1+4\pi\chi_{\alpha\alpha}$. As a result, Eq.~(\ref{08}) yields the following dispersion relation $k^x(\omega)$:
\begin{eqnarray}
\label{09}
k^x_{\pm}(\omega)=\pm k(\omega)=\pm\sqrt{\mu_{zz}}\frac{\omega}{c}.
\end{eqnarray}
Two signs in the r.h.s. of this equation correspond to for\-ward and backward propagation. Obviously, the velocity for traveling forward and backward is the same (directional reciprocity holds).
Similarly, for the case 2, 
\begin{eqnarray}
\label{10}
k^x_{\pm}(\omega)=\pm k(\omega)=\pm\sqrt{\mu_{yy}}\frac{\omega}{c}.
\end{eqnarray}

Clear, $k(\omega)$ in the dispersion relations (\ref{09}) and (\ref{10}) is, generally speaking, a complex function, i.e.,  $k(\omega)=k^{\prime}(\omega)+{\rm i}k^{\prime\prime}(\omega)$. The real part $k^{\prime}(\omega)$ yields the  wave's phase velocity $v_{\omega}=\omega/k^{\prime}(\omega)$, while the imaginary part $k^{\prime\prime}(\omega)>0$ forms the attenuation coefficient. Besides, the ratio between the speed of light $c$ and the phase velocity $v_{\omega}$ is known as the (real part of) refractive index, $n_{\omega}^{\prime} = c / v_{\omega}$. The refractive index is also a complex-valued function of $\omega$: $n_{\omega}^\prime +{\rm i} n_{\omega}^{\prime\prime}=c(k^\prime +{\rm i}k^{\prime\prime})/\omega$.

To summarize this section, the electromagnetic wave propagation in a magnetic medium, when  magnetic component of the electromagnetic wave weakly interacts with a set of quantum spin chains, has been examined within the frames of the classical electrodynamics of a continuous medium. The dispersion relation involves the dynamic (i.e., frequency-dependent) susceptibilities for the quantum-spin-chain medium under consideration. Therefore, we turn now to calculation of dynamic susceptibilities for free-fermion quantum spin chain (\ref{01}).

\section{Dynamic susceptibilities}
\label{s4}

To proceed the analysis of propagation of the plane electromagnetic wave in the magnetic crystal, defined in Sec.~\ref{s2}, we need the susceptibilities $\chi_{zz}({\bm k},\omega)$ and $\chi_{yy}({\bm k},\omega)$,  
defined in Eq.~(\ref{07}). After all, they enter the dispersion relation according to Eqs.~(\ref{09}) and (\ref{10}). 
On the other hand, we can calculate the susceptibilities $\chi_{\alpha\beta}(\kappa,\omega)$ for quantum spin model (\ref{01}),  
\begin{eqnarray}
\label{11}
\chi_{\alpha\beta}(\kappa,\omega){=}\frac{1}{m_0^2}\sum_{n=1}^Ne^{{\rm i}\kappa n}\int\limits_{0}^{\infty}{\rm d}t e^{{\rm i}\omega t}{\rm i}\langle\left[m_j^{\alpha}(t),m_{j+n}^{\beta}\right]\rangle,
\end{eqnarray}
at $\kappa{=}0$; 
$\chi_{\alpha\beta}(\kappa{=}0,\omega)$ (\ref{11}) should be proportional to $\chi_{\alpha\beta}({\bm k},\omega)$ defined in Eq.~(\ref{07}). Bearing in mind the inspection of frequency dependencies, we set for simplicity the constant of proportionality to $1$. As a consequence, we will illustrate how the electromagnetic-wave propagation can be controlled by varying the parameter ${\cal B}$ in Eq.~(\ref{01}). From here on out, ${\cal B}$ incorporates $m_0$, i.e., ${\cal B}m_0\to {\cal B}$ in Eq.~(\ref{01}).

More precisely, we need the dynamic (frequen\-cy-dependent) susceptibility
\begin{eqnarray}
\label{12}
\chi_{\alpha\alpha}(0,\omega){=}\sum_{n=1}^N\int\limits_{0}^{\infty}{\rm d}t e^{{\rm i}\omega t}{\rm i}\langle\left[s_j^{\alpha}(t),s_{j+n}^{\alpha}\right]\rangle
\end{eqnarray}
with $\alpha=z$ and $\alpha=y$. While $\chi_{zz}(\kappa,\omega)$ is governed by two-fermion excitation continua and was examined in depth analytically, see Ref.~\cite{Taylor1985}, $\chi_{xx}(\kappa,\omega)=\chi_{yy}(\kappa,\omega)$ is governed by many-fermion excitation continua and was examined in great detail numerically, see Refs.~\cite{Stolze1995,Derzhko1997,Derzhko1998,Derzhko2000,Derzhko2006,Krokhmalskii2008}.
Noting that $[\sum_{n=1}^Ns_n^z,H]=0$, one immediately concludes that $\chi_{zz}(0,\omega)=0$. In contrast, $\chi_{yy}(\kappa,\omega)=\chi_{xx}(\kappa,\omega)$ may be nonzero at $\kappa=0$. Therefore, for the spin-$1/2$ isotropic $XY$ chain in a transverse field in case 1 the dispersion relation remains as in vacuum, but in case 2 it alters.

\begin{figure}
	\includegraphics[width=\columnwidth]{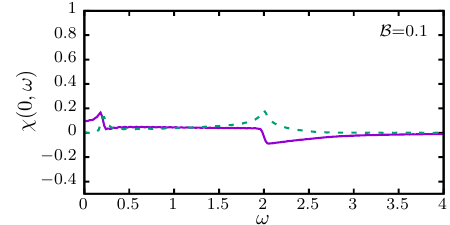}\\ 
	\includegraphics[width=\columnwidth]{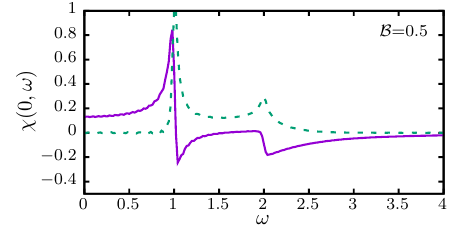}\\ 
	\includegraphics[width=\columnwidth]{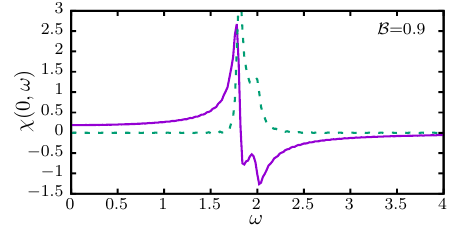}\\
	\includegraphics[width=\columnwidth]{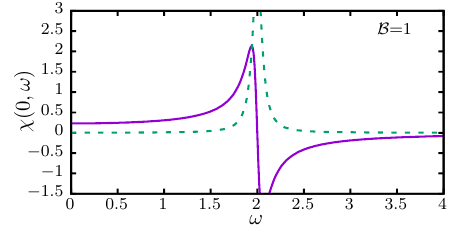}\\
	\includegraphics[width=\columnwidth]{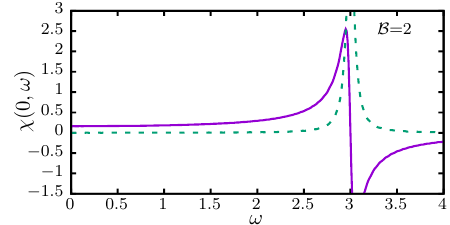}
	\caption{$\chi_{xx}^{\prime}(\kappa=0,\omega)$ (solid) and $\chi_{xx}^{\prime\prime}(\kappa=0,\omega)$ (dashed) for the spin-1/2 isotropic $XY$ chain in a transverse field. $J=1$, ${\cal B}$ incorporates $m_0$, $T=0.02$. From top to bottom: ${\cal B}=0.1,\,0.5,\,0.9,\,1$, and $2$.}
	\label{f2}
\end{figure}

We calculate $\chi_{xx}(0,\omega)$ in Eq.~(\ref{12}) numerically following the lines of Refs.~\cite{Derzhko1997,Derzhko1998,Derzhko2000} (for more recent study  see \cite{Khodaeva2026}) and report our findings in Fig.~\ref{f2}. 
In our calculations, we take an open chain of $N=1600$ sites and set $J{=}1$ (antiferromagnetic coupling) and ${\cal B}m_0\to{\cal B}$ in Eq.~(\ref{01}). After Jordan-Wigner fermionization (\ref{02}), we determine numerically the eigenvalues and eigenvectors of the tridiagonal symmetric matrix $(H_{nm})$. According to Eq.~(\ref{03}), 
$4\langle s_j^x(t)s_{j+n}^x\rangle
=\langle\varphi_1^+(t)\varphi_1^-(t){\ldots}
\varphi_{j{+}n{-}1}^+\varphi_{j{+}n{-}1}^-\varphi_{j+n}^+\rangle$,
i.e., we face a thermodynamic average of $2(2j+n+1)$ $\varphi^{\pm}$-operators, which are linear combinations of creation and annihilation operators, in terms of which the Hamiltonian is diagonal. In our calculations, we set $j=301$ and compute correlations up to $n=100$. Thus, we have to apply Wick's theorem for fermions with the result that can be compactly written as a Pfaffian of the antisymmetric matrix constructed from elementary contractions, i.e.,
\begin{eqnarray}
\label{13}
\langle\varphi_1{\ldots}\varphi_{2m}\rangle{=}{\rm pf}
\left(
\begin{array}{cccc}
0 & \langle\varphi_1\varphi_2\rangle & {\ldots} & \langle\varphi_1\varphi_{2m}\rangle \\
{-}\langle\varphi_1\varphi_2\rangle & 0 & {\ldots} & \langle\varphi_2\varphi_{2m}\rangle \\
\vdots & \vdots & \vdots & \vdots \\ 
{-}\langle\varphi_1\varphi_{2m}\rangle & {-}\langle\varphi_2\varphi_{2m}\rangle & {\ldots} & 0
\end{array}
\right).
\end{eqnarray} 
Four different elementary contractions 
$\langle\varphi^+_j(t)\varphi^+_m\rangle$,
$\langle\varphi^+_j(t)\varphi^-_m\rangle$,
$\langle\varphi^-_j(t)\varphi^+_m\rangle$,
and
$\langle\varphi^-_j(t)\varphi^-_m\rangle$
in Eq.~(\ref{13}) are the known functions of the eigenvalues and eigenvectors of the tridiagonal symmetric matrix  $(H_{nm})$ \cite{Derzhko1998}.
In this way we obtain $\langle s_j^x(t)s_{j+n}^x\rangle$ at time $t$.
Then we integrate over time in Eq.~(\ref{12}) with the step $\Delta t{=}0.2$ till $t=100$. 
We also introduce a small damping parameter changing the real $\omega$ to $\omega{+}{\rm i}\varepsilon$ with $\varepsilon=0$ (small ${\cal B}<1$) or $\varepsilon=0.05$ (large ${\cal B}\geq 1$). In addition, instead of zero temperature we consider sufficiently low one setting $T=0.02$. The real and imaginary parts of $\chi_{xx}(0,\omega)$ (\ref{12}) are reported in Fig.~\ref{f2}. 

In addition, we performed several cross-checkings of our calculations \cite{Baran2025}.
Namely, (i) we calculated the dynamic structure factor $S_{xx}(0,\omega)=\sum_{n=1}^N \int_{-\infty}^{\infty}{\rm d}te^{-\varepsilon\vert t\vert}e^{{\rm i}\omega t}\langle s_j^x(t)s_{j+n}^x\rangle$, which gives the imaginary part of $\chi_{xx}(0,\omega)$: $\chi^{\prime\prime}_{xx}(0,\omega)=(1-e^{-\omega/T})S_{xx}(0,\omega)/2$. Then, using the Kramers-Kronig relation, $\chi^{\prime}_{xx}(0,\omega){=}(1/\pi){\cal P}\int_{-\infty}^{\infty}{\rm d}\varpi \chi^{\prime\prime}_{xx}(\varpi)/(\varpi-\omega)$ (${\cal P}$ denotes the Cauchy principal value), we obtained the real part of $\chi_{xx}(0,\omega)$, too. The obtained results are consistent with the direct calculations of $\chi_{xx}(0,\omega)=\chi^{\prime}_{xx}(0,\omega)+{\rm i}\chi^{\prime\prime}_{xx}(0,\omega)$.
Next, (ii) in the strong-field zero-temperature limit the time-dependent correlation functions $\langle s_j^{x}(t)s_{j+n}^{x}\rangle$ can be calculated rigorously \cite{Cruz1981} (see also Refs.~\cite{Derzhko2002,Jedrzejewski2008,Derzhko2008}). As a result, we immediately obtain $\chi_{xx}(0,\omega)$ following the definition, cf. Eq.~(\ref{12}). The results are indistinguishable from the ones obtained by numerical calculation at $T=0.02$ in the panels with ${\cal B}=1,2$ of Fig.~\ref{f2}.
Finally, (iii) the strong-field zero-temperature limit case can be understood from a completely different perspective, that is, using the Landau-Lifshitz equation, which  describes the magnetization vector in ferromagnetic material, see Ref.~\cite{Landau1984}, page~270, problem 1. Again, the Landau-Lifshitz-equation prediction for $\chi_{xx}(0,\omega)$ is indistinguishable from what is reported in the panels with ${\cal B}=1,2$ of Fig.~\ref{f2}. For further details see Ref.~\cite{Baran2025}.

As said above, $\chi_{xx}(\kappa,\omega)$ is obviously related to $S_{xx}(\kappa,\omega)$; the latter quantity was discussed in great detail in Ref.~\cite{Derzhko2000}. Within the Jordan-Wigner picture, such dynamic quantities involve many-fermion excitations and therefore are not {\it a priori} restricted to certain regions in the $(\kappa,\omega)$ plane. However, as it was illustrated in Ref.~\cite{Derzhko2000}, they show, as a matter of fact, washed-out excitation brunches which follow the two-fermion excitation continuum boundaries. The two-fermion excitation continuum boundaries in the $(\kappa,\omega)$ plane are well known, see Refs.~\cite{Mueller1981,Derzhko2005}. Rethinking the analysis of Ref.~\cite{Derzhko2000} in the context of $\chi_{xx}(0,\omega)$ for the antiferromagnetic chain (1), one immediately concludes that, e.g., the drops/peaks of the solid/dashed curves at $\omega=2$ in the panels with ${\cal B}\le 1$ occur just at the upper boundary of the two-fermion excitation continuum.

\begin{figure}
	\includegraphics[width=\columnwidth]{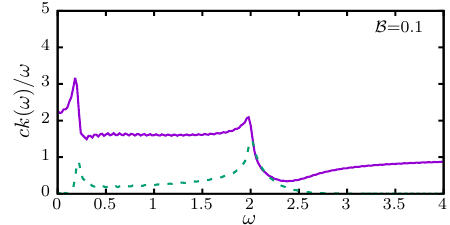}\\
	\includegraphics[width=\columnwidth]{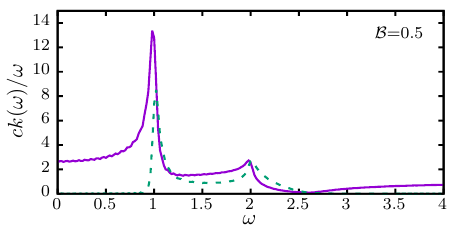}\\
	\includegraphics[width=\columnwidth]{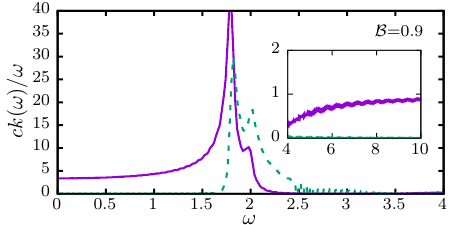}\\
	\includegraphics[width=\columnwidth]{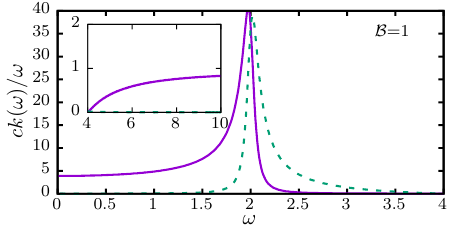}\\
	\includegraphics[width=\columnwidth]{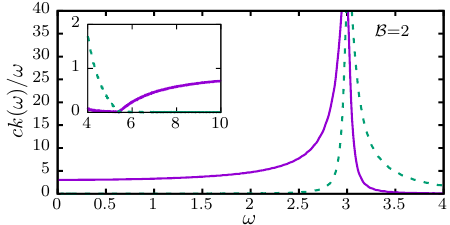}
	\caption{Towards the dispersion relation $k(\omega)$, see Eq.~(\ref{14}): $ck^{\prime}(\omega)/\omega$ (solid) and $ck^{\prime\prime}(\omega)/\omega$ (dashed) for the spin-1/2 isotropic $XY$ chain in a transverse field. $J=1$, ${\cal B}$ incorporates $m_0$, and $T=0.02$. From top to bottom: ${\cal B}=0.1,\,0.5,\,0.9,\,1$, and $2$. The insets present the same data as in the main panels but for higher frequencies.}
	\label{f3}
\end{figure}

Furthermore, we insert $\chi_{xx}(\kappa=0,\omega)$ into Eq.~(\ref{10}) to obtain for $k(\omega)=k^{\prime}(\omega)+{\rm i}k^{\prime\prime}(\omega)$
\begin{eqnarray}
	\label{14}
	\frac{ck(\omega)}{\omega}=\sqrt{1+4\pi\left(\chi_{xx}^{\prime}(0,\omega)+{\rm i}\chi_{xx}^{\prime\prime}(0,\omega)\right)}
	\nonumber\\
	=\sqrt{\sqrt{\left(1+4\pi \chi_{xx}^{\prime}(0,\omega)\right)^2+\left(4\pi \chi_{xx}^{\prime\prime}(0,\omega) \right)^2}} e^{{\rm i}\frac{\alpha}{2}},
	\nonumber\\
	\tan \alpha=\frac{4\pi \chi_{xx}^{\prime\prime}(0,\omega)}{1+4\pi \chi_{xx}^{\prime}(0,\omega)}.
\end{eqnarray}
The obtained frequency dependencies $k^{\prime}(\omega)$ and $k^{\prime\prime}(\omega)$ are reported in Fig.~\ref{f3}. We discuss these results in the next section. We note in closing that $ck^{\prime}(\omega)/\omega \ge 1$ means that the wave's phase velocity does not exceed $c$. (The phase velocity can be greater than $c$; this is not a violation of special relativity. However, the group velocity should be always less than or equal to $c$.)

\section{Discussion, summary, and outlook}
\label{s5}

Let us discuss the electromagnetic wave propagation through the magnetic medium consisting of noninteracting quantum spin chains, each of which with the Hamiltonian given in  Eq.~(\ref{01}), as it follows from the analysis of Secs.~\ref{s3} and \ref{s4}.  

There are two parameters in the problem: $\omega/J$ (electromagnetic wave frequency) and ${\cal B}/J$ (magnetic field magnitude). In the high-frequency limit $\omega/J\gg 1$, as can be seen in Fig.~\ref{f3}, $ck^{\prime}/\omega \to 1$, $ck^{\prime\prime}/\omega\to 0$ (also for ${\cal B}=0.9,\,1,\,2$, see the insets in Fig.~\ref{f3}) that indicates electromagnetic wave propagation as in vacuum: Electromagnetic wave of high frequency simply does not notice the quantum spin chain. In the opposite low-frequency limit $\omega/J\ll 1$, as can be seen in Fig.~\ref{f3}, $ck^{\prime}/\omega >ck^{\prime\prime}/\omega\to 0$ that illustrates again no attenuation of the low-frequency electromagnetic wave.

More complicated picture emerges at intermediate values of $\omega/J$ (terahertz frequency range) when attenuation shows up (dashed curves in Fig.~\ref{f3}). Moreover, the value of ${\cal B}/J$ is important, too. In the case of strong field ${\cal B}/J\ge 1$, one observes a diminishing of the electromagnetic wave velocity accompanied by damping at $\omega=J+{\cal B}$, see the panels for ${\cal B}=2,\,1$ in Fig.~\ref{f3}. (This is in accordance with the behavior of $\chi_{xx}(0,\omega)$ shown in Fig.~\ref{f2}.) Such a behavior of medium may resemble a noninteracting spin 1/2, which adsorbs the electromagnetic-wave energy. The case of weak field ${\cal B}/J< 1$ is even more intricate, see the panels for ${\cal B}=0.9,\,0.5,\,0.1$ in Fig.~\ref{f3}. Namely, according to these plots, there are two characteristic frequencies, $\omega_1=2{\cal B}$ and $\omega_2=2$. In the frequency range between $\omega_1$ and $\omega_2$, for ${\cal B}=0.1$ and ${\cal B}=0.5$, $ck^{\prime}(\omega)/\omega>ck^{\prime\prime}(\omega)/\omega$, both quantities are comparable that implies a propagation with reduced velocity and some damping. For ${\cal B}=0.9$, $\omega_1$ almost approaches $\omega_2$ and the picture resembles qualitatively the ones for ${\cal B}=1$, cf. the corresponding panels in Fig.~\ref{f3}. 
In Fig.~\ref{f3}, we observe rather wide frequency regions of the so-called anomalous dispersion where ${\rm d}n^{\prime}_{\omega}/{\rm d}\omega <0$ (for most of the frequency range ${\rm d}n^{\prime}_{\omega}/{\rm d}\omega >0$ (normal dispersion)). 
Obviously, by varying ${\cal B}$, one can control a spreading of electromagnetic wave through the media under consideration.

Lastly, the above discussion concerns $T\to 0$. For the high-temperature limit $T\to\infty$, only the autocorrelation $\langle s_j^{x}(t)s_{j}^{x}\rangle$ survives. Moreover, it follows a Gaussian decay as $t$ increases \cite{Brandt1976}. Therefore, $\chi_{xx}(0,\omega)$ (\ref{12}) vanishes in the high-temperature limit.

To summarize, we have presented a rigorous analysis of electromagnetic wave propagation through a quantum-spin-chain medium which is based on the macroscopic electrodynamics and the microscopic calculation of relevant dynamic susceptibilities for a free-fermion quantum spin chain. While the strong-field limit can be understood even within a phenomenological picture based on the Landau-Lifshitz equation, the opposite limit, when intersite interactions are relevant, requires a microscopic calculation of  the space- and time-dependent equilibrium correlations. Note that the bosonization method \cite{Giamarchi2003}, which exploits the bosonic nature of low-lying excitations and is often used to examine a spin-$1/2$ Heisenberg antiferromagnet in one dimension, is inapplicable here: Bosonization fails to describe high-frequency phenomena. 
Our research suggests that terahertz electromagnetic waves spreading along the chain direction may undergo noticeable changes in their dispersion relation ${\bm k}(\omega)$. Such changes can be controlled through an external field. From the solid-state point of view, our analysis may be useful as a guide while considering more realistic models which are not exactly solvable.

An interesting straightforward generalization of the present study would be a consideration of a magnetoelectric quantum spin chain with the so called Katsura-Nagaosa-Balatsky (KNB) mechanism for magnetoelectricity \cite{Katsura2005} (see also Refs.~\cite{Miyahara2016,Bolens2018a,Bolens2018b,Chari2021,Furuya2024,Solovyev2025}). The KNB scenario for emerging of magnetoelectricity has been incorporated into some exactly solvable quantum chain models \cite{Brockmann2013,Menchyshyn2015,Sznajd2018,Baran2018,Strecka2020,Richter2022,Brenig2025}. Preliminary analysis shows \cite{Baran2025} that switching on the KNB mechanism leads to the directional nonreciprocity, which implies a difference in the electromagnetic wave propagation between opposite directions along the chain. This feature may be related to a diode effect, when magnetoelectric system allows terahertz electromagnetic wave to propagate only in one direction (in analogy to the nonreciprocal resistive charge transport in semiconducting diode). A work in this direction is currently in progress.
\\
 


\noindent
{\bf {Acknowledgements:}} The authors are thankful to the Armed Forces of Ukraine for protection since 2014, and especially since February 24, 2022. 
O.~D. thanks the Abdus Salam International Centre for Theoretical Physics (Trieste) for kind hospitality at the Joint ICTP-WE Heraeus School and Workshop on Advances in Quantum Matter: Pushing the Boundaries, 4 -- 15 August 2025. O.~D. also thanks A.~Honecker, J.~Stre\v{c}ka, and K.~Karľov\'{a} for kind hospitality at the COOLMAG2025 Workshop Magnetic Cooling and Frustrated Magnetism, October 27th -- 29th, 2025, Ko\v{s}ice, Slovakia.

\noindent
{\bf {Research ethics:}} Not applicable.

\noindent
{\bf {Informed consent:}} Not applicable.

\noindent
{\bf {Author contributions:}} The authors have accepted responsibility for the entire content of this manuscript and approved its submission.

\noindent
{\bf {Use of Large Language Models, AI and Machine Learning Tools:}} None declared.

\noindent
{\bf {Conflict of interest:}} The authors state no conflict of interest.

\noindent
{\bf {Research funding:}} This study got funding within the IEEE program Magnetism for Ukraine 2025, supported by the IEEE Magnetic Society under the Science and Technology Center in Ukraine framework; project title: Dynamic properties of one-dimensional magnetoelectric crystal.

\noindent
{\bf {Data availability:}} Not applicable.

\bibliography{magnetoelectrics_pap}

\end{document}